\documentclass[twocolumn]{aastex701}

\usepackage{CJKutf8}
\usepackage{soul}
\usepackage{xcolor}

\sethlcolor{yellow}

\newcommand{\sersic}{S\'ersic}

\begin{document}
\title{LCEz4-M1: A Lyman Continuum Emitter Candidate at $z=4.444$ in the MUSE Hubble Ultra Deep Field}

\author[0000-0002-2528-0761]{Shuairu Zhu}
\affiliation{Shanghai Astronomical Observatory, Chinese Academy of Sciences, 80 Nandan Road, Shanghai 200030, People's Republic of China}
\affiliation{School of Astronomy and Space Sciences, University of Chinese Academy of Sciences, No. 19A Yuquan Road, Beijing 100049, People's Republic of China}
\email{shuairuz@shao.ac.cn}

\author[0000-0002-9634-2923]{Zhen-Ya Zheng}
\affiliation{Shanghai Astronomical Observatory, Chinese Academy of Sciences, 80 Nandan Road, Shanghai 200030, People's Republic of China}
\email[show]{zhengzy@shao.ac.cn}
\correspondingauthor{Zhen-Ya Zheng}

\author[0000-0002-1620-0897]{Fuyan Bian}
\affiliation{European Southern Observatory, Alonso de C\'ordova 3107, Casilla 19001, Vitacura, Santiago 19, Chile}
\affiliation{Chinese Academy of Sciences South America Center for Astronomy, National Astronomical Observatories, CAS, Beijing 100101, People's Republic of China.}
\email{Fuyan.Bian@eso.org}

\author[0000-0001-6763-5869]{Fang-Ting Yuan}
\affiliation{Shanghai Astronomical Observatory, Chinese Academy of Sciences, 80 Nandan Road, Shanghai 200030, People's Republic of China}
\email{yuanft@shao.ac.cn}

\author[0000-0002-0003-8557]{Chunyan Jiang}
\affiliation{Shanghai Astronomical Observatory, Chinese Academy of Sciences, 80 Nandan Road, Shanghai 200030, People's Republic of China}
\email{cyjiang@shao.ac.cn}

\author{Xiaer Zhang}
\affiliation{Shanghai Astronomical Observatory, Chinese Academy of Sciences, 80 Nandan Road, Shanghai 200030, People's Republic of China}
\email{zxe@shao.ac.cn}

\author[0000-0003-3987-0858]{Ruqiu Lin}
\affiliation{Department of Astronomy, University of Massachusetts, Amherst, MA 01003, USA}
\email{ruqiulin@umass.edu}

\author[0000-0002-6137-0422]{Yucheng Guo}
\affiliation{School of Earth \& Space Exploration, Arizona State University, 781 Terrace Mall, Tempe, AZ 85287, USA}
\email{yuchengg@asu.edu}

\begin{abstract}
High-redshift Lyman continuum emitters (LCEs) are crucial for understanding how galaxies ionize the neutral hydrogen in the epoch of reionization. However, detected LCEs at $z>4$ are quite rare. Here we report an LCE candidate at $z = 4.444$, dubbed LCEz4-M1, which is one of the highest-redshift LCE candidates currently known. The redshift is determined from the Ly$\alpha$ emission line detected in the VLT/MUSE spectrum. The Lyman continuum (LyC) signal is detected independently in the \emph{Hubble Space Telescope} (HST) F435W image and the VLT/MUSE spectrum at significances of $\simeq3.7~\sigma$ and $\simeq2.8-3.0~\sigma$, respectively. The LyC centroid is spatially consistent with the JWST/NIRCam continuum within the astrometric uncertainty. Adopting the maximum IGM transmission, we infer conservative lower-limit escape fractions of $f_{\rm esc}({\rm F435W}) = 0.82^{+0.13}_{-0.17}$ and $f_{\rm esc}({\rm MUSE}) = 0.75^{+0.18}_{-0.28}$. Using the combined JWST and MUSE data set, we characterize the physical properties and morphology of LCEz4-M1. In our fiducial JWST-only SED fit, the galaxy is compact but has a moderate current galaxy-integrated star formation surface density, $\Sigma_{\rm SFR}=0.38~M_{\odot}\,{\rm yr^{-1}\,kpc^{-2}}$, suggesting that it is not an extreme compact starburst under this fiducial interpretation.
While we find no clear evidence for an ongoing major merger for LCEz4-M1, the presence of a faint companion ($\sim 0\farcs5$) detected in the F277W band suggests a potential minor interaction. We also find that LCEz4-M1 may lie in a locally overdense region, although the environmental interpretation remains tentative.
Finally, the low ${\rm SFR}_{10\,{\rm Myr}}/{\rm SFR}_{100\,{\rm Myr}}$ ratio, low Ly$\alpha$ EW, and relatively weak rest-frame optical emission lines of LCEz4-M1 may indicate a post-burst LyC-leaking phase.
\end{abstract}


\section{Introduction}
Cosmic reionization is the final major phase transition of the universe, during which most hydrogen in the intergalactic medium (IGM) became ionized by Lyman-continuum (LyC) photons with $\lambda_\textrm{rest}<912$~\AA\ produced by astrophysical sources.
The timing of this transition is constrained by multiple observational probes.  CMB polarization measurements imply an integrated Thomson optical depth of $\tau = 0.054 \pm 0.007$, corresponding to an instantaneous reionization redshift of $z_{\rm re}\simeq7.7$ \citep{Planck2020}. Quasar damping-wing and Ly$\alpha$ emitter statistics indicate a substantially neutral IGM at $z\sim7$ \citep{Banados2018,Mason2018}, whereas quasar Gunn--Peterson and dark-pixel analyses suggest that reionization was largely complete by $z\sim6$ \citep{Fan2006,McGreer2015}.
Star-forming galaxies are now considered the primary contributors to the LyC photon budget required for reionization \citep[e.g.,][]{Robertson2022,Jiang2022,Jiang2025}.

Quantifying the ionizing photon contribution from galaxies is central to understanding cosmic reionization, but remains observationally challenging. The escape fraction of LyC photons into the IGM, $f_{\rm esc}$, is difficult to measure directly at $z\gtrsim4.5$ because the IGM becomes increasingly opaque to ionizing radiation. As a result, indirect constraints from lower-redshift Lyman Continuum Emitters (LCEs) have played a key role, providing laboratories for testing LyC escape mechanisms and informing expectations for galaxies in the reionization era.

Since the early 2000s, numerous ground-based searches have reported tentative LyC detections at $z \sim 3$ \citep[e.g.,][]{shapley2006, Iwata2009}. However, in the absence of high-resolution images, many early candidates were difficult to interpret and were likely affected by foreground contamination from low-redshift sources \citep{Vanzella2010, Grazian2016}. With continued observational efforts and, in particular, the availability of high-resolution imaging to identify and exclude interlopers, subsequent studies have established a small number of confirmed high-$z$ LCEs, while additional candidates remain to be verified \citep[See Table A1 of][]{Yuan2024}.

Despite extensive efforts, the common properties of LCEs at high redshift remain a matter of debate. Systematic surveys at $z\sim3$ suggest that galaxies with larger Ly$\alpha$ equivalent widths tend to exhibit higher LyC escape fractions \citep{Steidel2018, Fletcher2019}, yet other studies find no convincing LyC detections even among strong Ly$\alpha$ emitters \citep{Bian2020}, indicating that LyC escape may depend on additional factors such as viewing angle, transient ISM conditions, or sample selection effects.

In the nearby universe, where LyC detections are less affected by IGM absorption, more than 50 LCEs have been identified at low redshift \citep{Izotov2018a,Izotov2018b, Flury2022a, Roy2025}. However, these sources exhibit significant diversity in their physical properties. Most LCEs found by surveys such as the Low-Redshift Lyman Continuum Survey (LzLCS) are low-mass, compact galaxies with extreme emission-line properties \citep{Flury2022a, Izotov2018a, Izotov2018b}, while other studies have revealed an additional population of LCEs that are more massive, metal-rich, and show less extreme ionization conditions \citep{Borthakur2014, Wang2019, Roy2025}.

Moreover, comparisons between low-$z$ and high-$z$ LCEs also suggest possible systematic differences in their inferred properties. A systematic analysis of the physical properties of 23 LCEs in GOODS-S shows that a starburst is not a necessary condition for LyC escape in this sample \citep{zhu2024}. Further morphological analysis finds no clear correlation between the LyC escape fraction and compactness for these galaxies \citep{Zhu2025}. At low redshift, enhanced starburst activity is a common characteristic of many known LCEs, and a correlation between $f_{\rm esc}$ and compactness is often reported in low-$z$ samples \citep{Izotov2018a, Flury2022a}.

In summary, both low-$z$ and high-$z$ LCEs exhibit substantial diversity, and their inferred properties remain uncertain across cosmic time. These uncertainties likely arise from varying physical conditions, which highlight the need for larger LyC leaker samples. Observations at higher redshift are particularly valuable, as they probe LyC escape closer to the reionization epoch and offer more direct insights into the physical conditions of the early universe.

At $z>4$, systematic LyC searches are expected to yield very low success rates because the average IGM transmission short-ward of the Lyman limit is small, strongly suppressing the observability of LyC emission. Nevertheless, LyC leakage can still be detected along unusually transparent sight lines and in extreme systems. For example, the bright galaxy \emph{Ion3} at $z\simeq4.0$ shows copious LyC leakage \citep[e.g.,][]{Vanzella2018}.
Additionally, a transient LyC emission  was detected from a $z \sim 4.8$ event by the Einstein Probe \citep{Levan2025}, likely originating from a gamma-ray burst afterglow. These rare high-redshift LCEs play an outsized role because they provide direct anchor points for the escape process under physical conditions closer to the end of reionization.

In this Letter, we report a candidate LCE at $z_{\rm Ly\alpha}=4.444$, dubbed LCEz4-M1, selected from the MUSE-HUDF survey \citep{Bacon2023}, making it one of the highest-redshift LCE candidates currently known. An independent analysis of the same source, referred to as MXDFz4.4, was recently presented by \citet{Goovaerts2026}. We describe the data used to assess the LyC emission and characterize the source in Section~\ref{sec:data}. We present redshift determination and LyC detection in Section~\ref{sec:lyc}, and we discuss the physical and star-forming properties in Section~\ref{sec:prop}. Throughout this paper, we adopt a flat $\Lambda$CDM cosmology with $\Omega_{M}=0.3$, $\Omega_{\Lambda}=0.7$, and $H_{0}=70~{\rm km~s^{-1}~Mpc^{-1}}$. Distances in kpc refer to the proper (physical) units. All magnitudes are reported in the AB system.

\section{Data}
\label{sec:data}
LCEz4-M1 is located in the Hubble Ultra Deep Field (HUDF), which has been covered by multiple surveys carried out with \emph{James Webb Space Telescope} (JWST), \emph{Hubble Space Telescope} (HST) , and Multi Unit Spectroscopic Explorer (MUSE) on the Very Large Telescope (VLT). We use these extensive datasets to validate its redshift and analyze the properties of this source. The coordinates of LCEz4-M1 are R.A. = 53.1577359 and Decl. = $-27.7822654$ (J2000).

Spectroscopic redshifts are crucial for identifying LyC emission at high-$z$. LCEz4-M1 is covered by observations from the MUSE-HUDF survey \citep{Bacon2017, Bacon2023}, which consists of a $3 \times 3$~arcmin$^2$ mosaic with 10~hr exposure, a $1 \times 1$~arcmin$^2$ deep field with 31~hr exposure, and the 141~hr adaptive-optics-assisted MUSE eXtremely Deep Field (MXDF; 1~arcmin diameter). LCEz4-M1 is listed in the survey catalog as a Lyman Alpha Emitter (LAE) at $z = 4.444$. We use the released data products for this source, including the extracted spectrum and data cube, to analyze the Ly$\alpha$ emission properties and further verify the redshift. LCEz4-M1 also falls within the footprint of JWST spectroscopic surveys such as the JWST Advanced Deep Extragalactic Survey \citep[JADES;][]{Eisenstein2023} and the First Reionization Epoch Spectroscopically Complete Observations \citep[FRESCO;][]{Oesch2023}, but has not been targeted by JADES NIRSpec observations nor detected in FRESCO slitless spectroscopy. We note that the non-detection in FRESCO is expected, because the F444W grism used during the FRESCO survey covers only rest-frame wavelengths of $\sim7000$-$9000$ \AA~at $z=4.444$, where no strong nebular emission lines are expected. We have also searched the archival data from ground-based telescopes such as VLT, but have found no additional observations.

To analyze the physical properties, escape fraction, and morphology of LCEz4-M1, we also use imaging and photometric data from JWST and HST. We use JWST images released by the JADES survey Data Release 2 \citep[JADES DR2, ][]{Eisenstein2023, Rieke2023} for morphological analysis and SED fitting, and HST images released by the Hubble Legacy Field \citep[HLF, ][]{Illingworth2016, Whitaker2019}, particularly HST/ACS WFC F435W and F814W, for estimating the escape fraction. We also use VLT/VIMOS $U$-band imaging to constrain the LyC emission of LCEz4-M1 \citep{nonino2009}. The JADES DR2 is constructed from JWST images from surveys including JADES \citep{Eisenstein2023}, JEMS \citep{Williams2023}, and FRESCO \citep{Oesch2023}. 

\section{The Redshift and LyC Signals}
\label{sec:lyc}
\subsection{Redshift Validation}
\begin{figure*}
    \centering
    \includegraphics[width=1\linewidth]{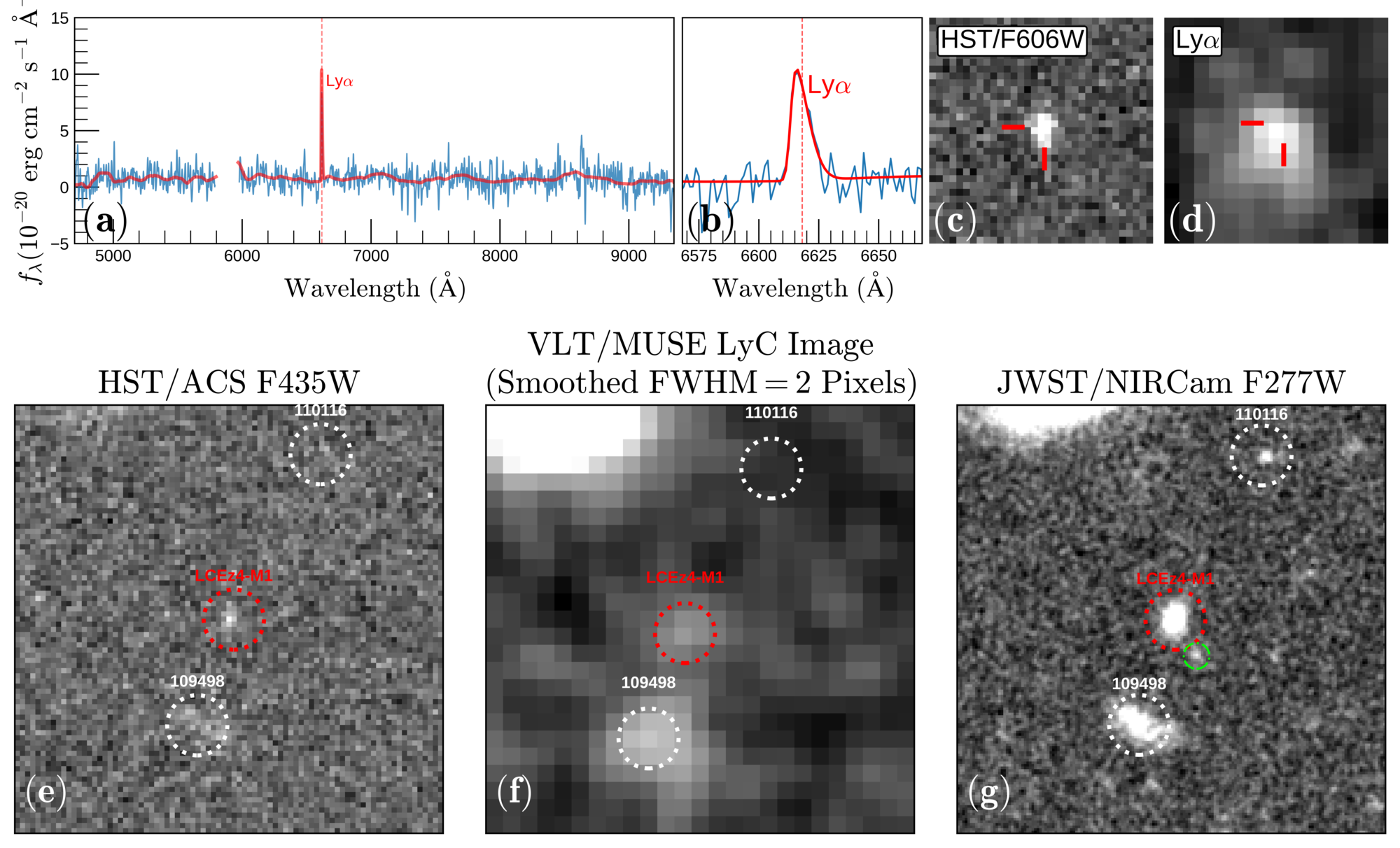}
    \caption{MUSE spectrum and multi-band cutouts of LCEz4-M1.
    (a) The full MUSE spectrum. No emission lines other than Ly$\alpha$ are detected.
    (b) The Ly$\alpha$ emission line in the MUSE spectrum.
    (c) The HST ACS/WFC F606W image, whose bandpass covers the Ly$\alpha$ line.
    (d) The Ly$\alpha$ narrowband image extracted from the MUSE data cube.
    (e) The HST ACS/WFC F435W cutout tracing the LyC emission.
    (f) The VLT/MUSE LyC narrowband image, smoothed with a FWHM of two pixels for display.
    (g) The JWST/NIRCam F277W cutout, probing the rest-frame optical continuum and the [O~\textsc{iii}] emission line. The sizes of each cutouts are $2\arcsec\times2\arcsec$ in panels (c) and (d), and $5\arcsec\times5\arcsec$ in panels (e)--(g). The red dotted circles ($r=0\arcsec.35$) mark LCEz4-M1, the white dotted circles ($r=0\arcsec.35$) mark nearby foreground sources, and the small green circle ($r=0\arcsec.15$) in panel (g) marks the companion candidate.}
    \label{fig:spec-z}
\end{figure*}

A reliable redshift is crucial for identifying LyC emission from high-$z$ galaxies. LCEz4-M1 falls within the footprint of the MUSE-MXDF observations from the MUSE-HUDF survey, with a total integration time of $\sim 140$~hr. The spectrum has a resolution of $R \sim 1800$ and was fitted by \citet{Bacon2023} using \texttt{pyPlatefit}. In the original fit, three sets of emission lines were detected at observer-frame 6620~\AA, 7600~\AA, and 8450~\AA, respectively. These detections correspond to Ly$\alpha$, Si~{\sc iv}~$\lambda1396.92$ + O~{\sc iv}]~$\lambda1397.23$, and the C~{\sc iv} doublet at $z = 4.444$. We examine the fitting results and find that Si~{\sc iv}~$\lambda1396.92$ + O~{\sc iv}]~$\lambda1397.23$ and the C~{\sc iv} doublet are detected at only $\sim 2~\sigma$ with velocity dispersions of $\sim 10~\rm km~s^{-1}$, suggesting that they are likely false positives. The emission line at 6620~\AA\ is detected at $\sim 11~\sigma$ and is highly asymmetric, consistent with Ly$\alpha$ profiles at high redshift (see Figure~\ref{fig:spec-z}b).

We note that LCEz4-M1 is listed in the JADES DR2 catalog as ID~124950, with a photometric redshift of $z_{\rm phot}=0.42$ \citep{Hainline2024}. This photometric redshift was derived from the combined JADES photometry and HST/ACS photometry from the HLF catalog. For this source, however, we find a systematic offset between the HLF ACS fluxes and the JADES photometry, with the ACS fluxes being systematically lower, which artificially reddens the UV continuum and likely drives the preference for the low-redshift solution. Nevertheless, we explicitly consider the possibility that the emission line detected at 6620~\AA\ is not Ly$\alpha$.

If the 6620~\AA\ line is physically associated with LCEz4-M1, one possible low-redshift interpretation is that it corresponds to the [O~{\sc iii}] $\lambda\lambda4959,5007$ doublet, implying $z \simeq 0.32$ (with the observed feature identified as [O~{\sc iii}] $\lambda5007$). In this case, the [O~{\sc iii}] $\lambda5007$ line is detected at $\sim 11\sigma$, and the [O~{\sc iii}] $\lambda4959$ component should also be detected at the expected wavelength at a significance of $\sim 4\sigma$. However, we do not find any significant emission at the expected position of [O~{\sc iii}] $\lambda4959$, with signal-to-noise ratio $(\rm S/N) <1$, which disfavors this low-redshift interpretation.

Another possible interpretation is that the detected feature is the [O~{\sc ii}] $\lambda\lambda3726,3729$ doublet, which would place LCEz4-M1 at $z \sim 0.78$. In this scenario, the apparent asymmetry of the line profile could arise from the blended [O~{\sc ii}] doublet. We therefore fit the emission line using both a double-Gaussian model (for [O~{\sc ii}]) and an asymmetric Gaussian model (for Ly$\alpha$), and find that the asymmetric Gaussian provides a better fit to the observed profile. We further examine the expected location of [O~{\sc iii}] under the [O~{\sc ii}] interpretation and find no significant emission at the corresponding wavelength, with ${\rm S/N} \sim 0$. Taken together, the line-profile fitting and the non-detection of [O~{\sc iii}] disfavour the [O~{\sc ii}] scenario.

Therefore, if the emission line originates from LCEz4-M1 itself, the available spectroscopic evidence favors a Ly$\alpha$ identification over the low-redshift [O~{\sc iii}] or [O~{\sc ii}] alternatives, meaning $z = 4.444$.

LCEz4-M1 has a companion separated by $\sim 0''.5$. Aperture photometry shows the companion is detected only in F277W (${\rm S/N}\sim 16$) and F356W (${\rm S/N} \sim 11$), consistent with [O~{\sc iii}] and H$\alpha$ emission at $z = 4.444$, suggesting an emission-line-dominated galaxy at the same redshift as LCEz4-M1. We measure the flux density ratio between the two sources and find that LCEz4-M1 dominates, with ratios of 7:1 and 8:1 in F277W and F356W, respectively. Although the MUSE-MXDF data do not resolve the pair, this large flux contrast indicates that the Ly$\alpha$ emission in the MUSE spectrum is dominated by LCEz4-M1.

Furthermore, if the Ly$\alpha$ emission originates from the companion rather than LCEz4-M1, this spatial offset should manifest as a centroid shift between the image of the narrowband and that of the broadband continuum, which covers the narrowband wavelength range. To quantify the expected offset, we simulate the two-source configuration using \texttt{GalSim} \citep{Rowe2015}, adopting a Moffat PSF with $\beta = 2.5$ and Full Width at Half Maximum (FWHM) $= 0''.55$ \citep[see][]{Bacon2023}, matched to the MUSE observations. The simulation predicts a centroid offset of $\sim 2.5$ MUSE pixels ($\sim 0''.5$) if the companion dominates the Ly$\alpha$ flux.

Comparing the HST/ACS F606W image with the Ly$\alpha$ narrowband image (Figure~\ref{fig:spec-z}c and d), we find a negligible offset between the centroids. To further verify this result using a consistent PSF, we construct two images directly from the MUSE data cube using the F606W transmission curve: (1) a continuum-only image, created by excluding the wavelength range containing the Ly$\alpha$ emission, and (2) a line-only image, obtained by subtracting the continuum image from the full integrated image. The measured centroid offset between these two images is $< 1$ MUSE pixel, confirming that the Ly$\alpha$ emission is spatially coincident with the UV continuum of LCEz4-M1.

Based on these two independent checks, we conclude that the Ly$\alpha$ emission originates primarily from LCEz4-M1, which confirms the source redshift at $z = 4.444$.

\subsection{\texorpdfstring{LyC Signals at $z=4.444$}{LyC Signals at z=4.444}}
LCEz4-M1 is an LAE at $z = 4.444$. At this redshift, the Lyman limit is redshifted to $\sim 4960$~\AA\ in the observer frame, meaning the Lyman continuum can be probed by HST imaging, including HST ACS/WFC F435W, Wide Field Camera 3 (WFC3)/UVIS F336W, F275W, and F225W images. In addition, part of the LyC spectrum falls within the wavelength coverage of VLT instruments such as the VIMOS U-band and MUSE. We constrain the LyC emission of LCEz4-M1 using these datasets.

We first measure LyC emission using the HST/ACS F435W image from the HLF, which covers a rest-frame wavelength range of 660-900~\AA. We perform aperture photometry on LCEz4-M1 using \texttt{photutils} with an aperture diameter of 0\farcs7, obtaining an HST/ACS F435W magnitude of 29.3~mag ($f_{\lambda} = 1.06 \times 10^{-20}$~erg~s$^{-1}$~cm$^{-2}$~\AA$^{-1}$, see Figure~\ref{fig:spec-z}e). We estimate the photometric uncertainty empirically by randomly placing 1000 apertures of the same size in blank regions near the source and using the resulting flux distribution as the local noise estimate. Comparing the F435W flux measured at the position of LCEz4-M1 with this distribution gives a detection significance of $\sim 3.7~\sigma$ and a $3~\sigma$ limiting depth of $\sim$29.8~mag in the vicinity of this source. We also examined the HLF epoch data in F435W and the other available bands. In F435W, LCEz4-M1 is covered by three epochs with non-zero exposure at the source position. These three epochs give an inverse-variance weighted magnitude of $m_{\rm AB}=29.39\pm0.29$ mag, corresponding to ${\rm S/N}=3.69$. Based on the HLF epoch data, we find no significant variability in F435W or in the other available bands.
This source is also listed in the HLF catalog. The HST/ACS F435W magnitude of LCEz4-M1 is 29.56~mag with $\rm S/N=3.2~\sigma$.

Based on the LyC emission detected in HST/ACS F435W, we measure the spatial offset between the LyC signal and the non-ionizing UV continuum, using JWST/NIRCam F200W as the reference. The measured offset is only $0\farcs06$. However, this value is comparable to the relative astrometric uncertainty between the HST F435W and JWST images, for which nearby matched sources show a median two-dimensional offset of $\sim0\farcs04$ and a scatter of $\sim0\farcs07$. We therefore do not regard this offset as significant. The F435W/LyC emission is spatially consistent with the F200W morphology within the astrometric uncertainty.

We also examine LyC emission using data from the MUSE-HUDF survey, which covers the LyC spectrum of LCEz4-M1 from rest-frame $864$~\AA\ to 912~\AA. We measure LyC emission in two independent ways, by integrating the flux in the Lyman continuum regime directly from the extracted spectrum of \citet{Bacon2023} and by constructing a narrowband image from the MUSE data cube covering the LyC wavelength range with aperture photometry using a diameter of 0\farcs7. Both methods yield consistent results, with the LyC emission detected at $\sim 2.8-3.0~\sigma$ significance (see Figure~\ref{fig:spec-z}f). The detection of the Lyman continuum signal through the F435W image and the MUSE data significantly enhances the reliability of the measurement, as it is unlikely to be mimicked by artifacts in both datasets.

However, the measured LyC flux densities from MUSE are lower than those derived from F435W photometry. This discrepancy may be attributed to aperture losses, as the 0\farcs7 diameter is smaller than the MUSE PSF FWHM, potentially missing extended flux. For the spectroscopic data, contamination from the nearby source (HLF ID: 109498) during spectral extraction from the data cube may affect the flux measurement.

We also perform an empirical false-positive test using spectroscopically confirmed LAEs at $z=4.0$--5.0 selected from the same MUSE HUDF DR2 and HST datasets. After matching to the JADES NIRCam catalog, restricting the sample to sources with F200W magnitudes within $\pm0.5$ mag of LCEz4-M1, and excluding objects with visually identified nearby contamination, the clean control sample contains 29 LAEs.

For the five clean LAEs with both MUSE and HST/F435W coverage, no random aperture among 500 trials simultaneously reaches the observed thresholds of ${\rm S/N}\geq3.7$ in F435W and ${\rm S/N}\geq3.0$ in the MUSE LyC image. The marginal false-positive rates are 0.0\% for F435W and 6.2\% for MUSE. For the additional clean LAEs with F435W coverage only, the F435W false-positive rate is 0.77\% among 2400 trials. Assuming independence between the two measurements, this implies a joint false-positive probability of only 0.0479\%.
Moreover, \citet{Goovaerts2026} reported a much higher F435W detection significance of $10.3\sigma$ for the same source using reprocessed HST/ACS F435W data from HST archival program PID 16621 (PI: Koekemoer).
The higher significance reflects the improved calibration and image processing of these data, including updated darks, biases, flatfields, and astrometric alignment, as well as other low-level improvements such as improved removal of cosmic rays, satellite trails, and bad pixels \citep{Goovaerts2026}. In addition, \citet{Goovaerts2026} adopted a UV-optimized photometric approach similar to that used by the previous studies \citep{Rafelski2015, Sun2024}, which would have larger SNR values than the fixed blank aperture method. Our measured LyC flux density is basically consistent with that reported by \citet{Goovaerts2026}, which are 6.7 $\pm$ 1.8 nJy and 4.2 $\pm$ 0.5 nJy, respectively. This independent high-S/N detection further disfavors the possibility that the LyC signal is a random fluctuation or a spurious source.

In addition to the F435W and MUSE detections, we also use HST/WFC3 UVIS imaging in F225W, F275W, and F336W, as well as VLT/VIMOS $U$-band data to constrain the Lyman continuum emission of LCEz4-M1. These filters probe wavelengths bluer than F435W. The WFC3 images are relatively shallow compared to F435W, reaching $5~\sigma$ depths of only $\sim 26.6$--$27.2$~mag near LCEz4-M1, while the VIMOS $U$-band is deeper at $\sim 29.4$~mag. However, no significant signal coincident with LCEz4-M1 is detected in any of these bands.

\begin{deluxetable}{l c}[th]
\tablecaption{Summary of measured and inferred properties for LCEz4-M1.\label{tab: res}}
\tablewidth{0pt}
\tabletypesize{\scriptsize}
\setlength{\tabcolsep}{6pt}
\tablehead{
\colhead{Property} & \colhead{Value}
}
\startdata
\multicolumn{2}{c}{\textbf{Ly$\alpha$ Properties}} \\
\tableline
$F(\mathrm{Ly}\alpha)^{\mathrm{a}}$ & $9.05 \pm 0.79$ \\
$L(\mathrm{Ly}\alpha)^{\mathrm{b}}$ & $0.179 \pm 0.016$ \\
$\mathrm{EW}(\mathrm{Ly}\alpha)$ (MUSE catalog; \AA) & $18.41 \pm 3.14$ \\
$\mathrm{EW}(\mathrm{Ly}\alpha)$ (HST-only; \AA) & $11.9 \pm 1.1$ \\
$\mathrm{EW}(\mathrm{Ly}\alpha)$ (JWST-only; \AA) & $6.5 \pm 0.6$ \\
S/N (Ly$\alpha$) & $11.46$ \\
\tableline
\multicolumn{2}{c}{\textbf{LyC Properties}} \\
\tableline
$f_{\lambda}(\mathrm{LyC})$ (F435W)$^{\mathrm{c}}$ & $1.06 \pm 0.29$ \\
$L_{\lambda}(\mathrm{LyC})$ (F435W)$^{\mathrm{d}}$ & $1.14 \pm 0.31$ \\
$f_{\lambda}(\mathrm{LyC})$ (MUSE)$^{\mathrm{c}}$ & $0.64 \pm 0.22$ \\
$L_{\lambda}(\mathrm{LyC})$ (MUSE)$^{\mathrm{d}}$ & $0.68 \pm 0.23$ \\
$f_{\rm esc}$ (F435W, MC) & $0.82^{+0.13}_{-0.17}$ \\
$f_{\rm esc}$ (MUSE, MC) & $0.75^{+0.18}_{-0.28}$ \\
$f_{\rm esc}$ (\textsc{CIGALE}) & $0.56 \pm 0.15$ \\
\tableline
\multicolumn{2}{c}{\textbf{Physical Properties (JWST-only)}} \\
\tableline
Age (Myr) & $45.93 \pm 16.08$ \\
$Z$ & $0.0001 \pm 0.0005$ \\
$\beta$ & $-2.26 \pm 0.13$ \\
$M_{\rm UV}$ (mag) & $-18.68$ \\
$\log(M_{\star}/M_{\odot})$ (dex) & $8.20 \pm 0.23$ \\
$\log({\rm SFR}/M_{\odot}\,{\rm yr^{-1}})$ (dex) & $-0.02 \pm 0.68$ \\
$\Sigma_{\rm SFR}$ ($M_{\odot}\,{\rm yr^{-1}\,kpc^{-2}}$) & $0.38$ \\
$E(B-V)$ (mag) & $0.0489 \pm 0.0132$ \\
${\rm SFR}_{10\,{\rm Myr}}$ ($M_{\odot}\,{\rm yr^{-1}}$) & $0.74 \pm 0.70$ \\
${\rm SFR}_{100\,{\rm Myr}}$ ($M_{\odot}\,{\rm yr^{-1}}$) & $4.67 \pm 1.49$ \\
\tableline
\multicolumn{2}{c}{\textbf{Physical Properties (HST-only)}} \\
\tableline
Age (Myr) & $5.93 \pm 5.67$ \\
$Z$ & $0.0012 \pm 0.0018$ \\
$\beta$ & $-1.44 \pm 0.18$ \\
$M_{\rm UV}$ (mag) & $-18.21$ \\
$\log(M_{\star}/M_{\odot})$ (dex) & $7.72 \pm 0.20$ \\
$\log({\rm SFR}/M_{\odot}\,{\rm yr^{-1}})$ (dex) & $1.22 \pm 0.40$ \\
$\Sigma_{\rm SFR}$ ($M_{\odot}\,{\rm yr^{-1}\,kpc^{-2}}$) & $6.59$ \\
$E(B-V)$ (mag) & $0.1861 \pm 0.0346$ \\
${\rm SFR}_{10\,{\rm Myr}}$ ($M_{\odot}\,{\rm yr^{-1}}$) & $12.38 \pm 9.31$ \\
${\rm SFR}_{100\,{\rm Myr}}$ ($M_{\odot}\,{\rm yr^{-1}}$) & $12.38 \pm 9.31$ \\
\tableline
\multicolumn{2}{c}{\textbf{Morphology}} \\
\tableline
$r_{50}$ (kpc) & $0.633 \pm 0.017$ \\
$\chi^2_{\nu}$ & $0.5$ \\
\tableline
\enddata
\tablenotetext{a}{Fluxes are in units of $10^{-19}\ \mathrm{erg\ s^{-1}\ cm^{-2}}$.}
\tablenotetext{b}{Luminosities are in units of $10^{42}\ \mathrm{erg\ s^{-1}}$.}
\tablenotetext{c}{LyC flux densities are in units of $10^{-20}\ {\rm erg\ s^{-1}\ cm^{-2}}$ \AA$^{-1}$.}
\tablenotetext{d}{$L_{\lambda}$ are in units of $10^{40}\ {\rm erg\ s^{-1}}$ \AA$^{-1}$.}
\tablenotetext{}{Uncertainties are quoted at $1~\sigma$ unless noted otherwise. The MC escape fractions are conservative lower-limit estimates computed with the maximum adopted IGM transmission. The physical properties are reported separately for the JWST-only and HST-only SED fits.}
\end{deluxetable}

\section{The Properties of LCEz4-M1}
\label{sec:prop}
In this section, we present an analysis of the properties of LCEz4-M1, including the LyC photon escape fraction, physical properties, as well as morphology and environments. All results are also summarized in Table~\ref{tab: res}.

\subsection{\texorpdfstring{Ly$\alpha$ and LyC Properties}{Ly-alpha and LyC Properties}}

The Ly$\alpha$ line does not exhibit complex features such as multiple peaks or broad wings (see Figure~\ref{fig:spec-z}b). The MUSE catalog gives a Ly$\alpha$ flux of $(9.05\pm0.79)\times10^{-19}~{\rm erg~cm^{-2}~s^{-1}}$ and a rest-frame emission equivalent width of $\mathrm{EW}(\mathrm{Ly}\alpha)=18.41\pm3.14$~\AA. For the SED-based estimates, we compute the continuum level at $1216$~\AA\ by extrapolating the SED-inferred $1500$~\AA\ continuum using the best-fit UV slope $\beta$ from each photometric solution. This gives $\mathrm{EW}(\mathrm{Ly}\alpha)=11.9\pm1.1$~\AA\ for the HST-only fit and $6.5\pm0.6$~\AA\ for the JWST-only fit (see Section~\ref{sec:sed}). The Ly$\alpha$ properties are summarized in Table~\ref{tab: res}.

We also estimate the escape fraction of LCEz4-M1, following the definition of \citet{Grazian2016, Siana2007, Steidel2001}:
\begin{equation}
    f_{\rm esc} = \frac{(L_{1500}/L_{\rm LyC})}{(f_{1500}/f_{\rm LyC})} \times \frac{1}{T_{\rm IGM}} \times e^{-\tau_{\rm UV,dust}}
\end{equation}
where $(f_{1500}/f_{\rm LyC})$ is the observed flux density ratio between 1500~\AA\ and the LyC regime, and $(L_{1500}/L_{\rm LyC})$ is the intrinsic luminosity ratio derived from stellar population models. The term $T_{\rm IGM}$ represents the IGM transmission at $z = 4.444$. The factor $e^{-\tau_{\rm UV,dust}}$ accounts for dust attenuation of the non-ionizing UV continuum.

We estimate the LyC escape fraction, $f_{\rm esc}$, using a Monte Carlo (MC) approach based on two independent observational datasets: HST photometry and MUSE spectroscopy.

For the photometric analysis, the observed flux density ratio $(f_{1500}/f_{\rm LyC})_{\rm obs}$ is determined from the HST F814W and F435W bands, which sample the rest-frame $\simeq 1480$~\AA\ and $\simeq 790$~\AA\ at $z=4.444$, respectively. The F435W and F814W flux densities are measured directly from the HLF images using the same $0\farcs35$-radius aperture. For the spectroscopic analysis, we derive the flux densities by integrating the MUSE spectrum over the LyC region and a 100~\AA\ window centered at rest-frame 1500~\AA.

The intrinsic luminosity-density ratio $(L_{1500}/L_{\rm LyC})_{\rm int}$ is derived from the fiducial JWST-only CIGALE mock SEDs generated by the posterior and fitted parameters. For each mock SED, we compute a proxy-matched intrinsic ratio using the relevant observational bandpass or wavelength window: $L_{\rm F814W}/L_{\rm F435W}$ for the HST-based estimate and $L_{\rm MUSE,1500}/L_{\rm LyC,MUSE}$ for the MUSE-based estimate. This procedure accounts for the finite filter throughputs and for the fact that the LyC measurements probe broad wavelength ranges rather than monochromatic 900~\AA\ flux densities. The dust attenuation term $e^{-\tau_{\rm UV,dust}}$ is computed from the corresponding fiducial JWST-only CIGALE dust parameters at 1500~\AA. We adopt a fixed IGM transmission of $T_{\rm IGM}=0.3334$, corresponding to the maximum transmission allowed by the \citet{Steidel2018} prescription at $z\simeq4.5$. This choice yields the most conservative estimate of the escape fraction. The inferred values should therefore be interpreted as lower limits under the fiducial JWST-only stellar-population and dust constraints.

We propagate the observed flux-density uncertainties by drawing positive fluxes from normal distributions centered on the measured F435W, F814W, MUSE LyC, and MUSE 1500~\AA\ flux densities. Combining these observed-flux draws with the fiducial JWST-only CIGALE mock SED posterior gives $10^4$ MC estimates for each LyC measurement. Even with this maximum IGM transmission, many mock SED realizations imply
$f_{\rm esc}>1$ for the observed LyC-to-UV flux ratios, indicating that the raw MC distribution should not be interpreted as a precise posterior on $f_{\rm esc}$. We therefore report the median and 16th--84th percentile range of the physically allowed subset with $f_{\rm esc}<1$, which contains 2.52\% and 5.53\% of the F435W- and MUSE-based MC samples, respectively. With this convention, we obtain $f_{\rm esc}({\rm F435W})=0.82^{+0.13}_{-0.17}$ and $f_{\rm esc}({\rm MUSE})=0.75^{+0.18}_{-0.28}$.
Note that \citet{Goovaerts2026} also reported a high F435W LyC detection and inferred a high escape fraction of roughly 50--100\% of the same object through a contemporaneous independent analysis, but with different stellar-population assumptions and IGM-transmission modeling.

In \citet{Zhu2025}, we compiled a sample of LCEs at $z>3$. The reported escape fractions span from $0.05$ to $0.88$, with a mean value of $0.41$. Using our F435W- and MUSE-based MC estimates, LCEz4-M1 lies toward the high-$f_{\rm esc}$ end of the sample.

\subsection{SED Fitting and Physical Properties}
\label{sec:sed}
We use Code Investigating GALaxy Emission \citep[\texttt{CIGALE} version 2025.1,][]{Burgarella2005, Noll2009, Boquien2019} to estimate the physical properties of LCEz4-M1 by fitting its SED.

\begin{figure*}[t]
    \centering
    \includegraphics[width=0.95\linewidth]{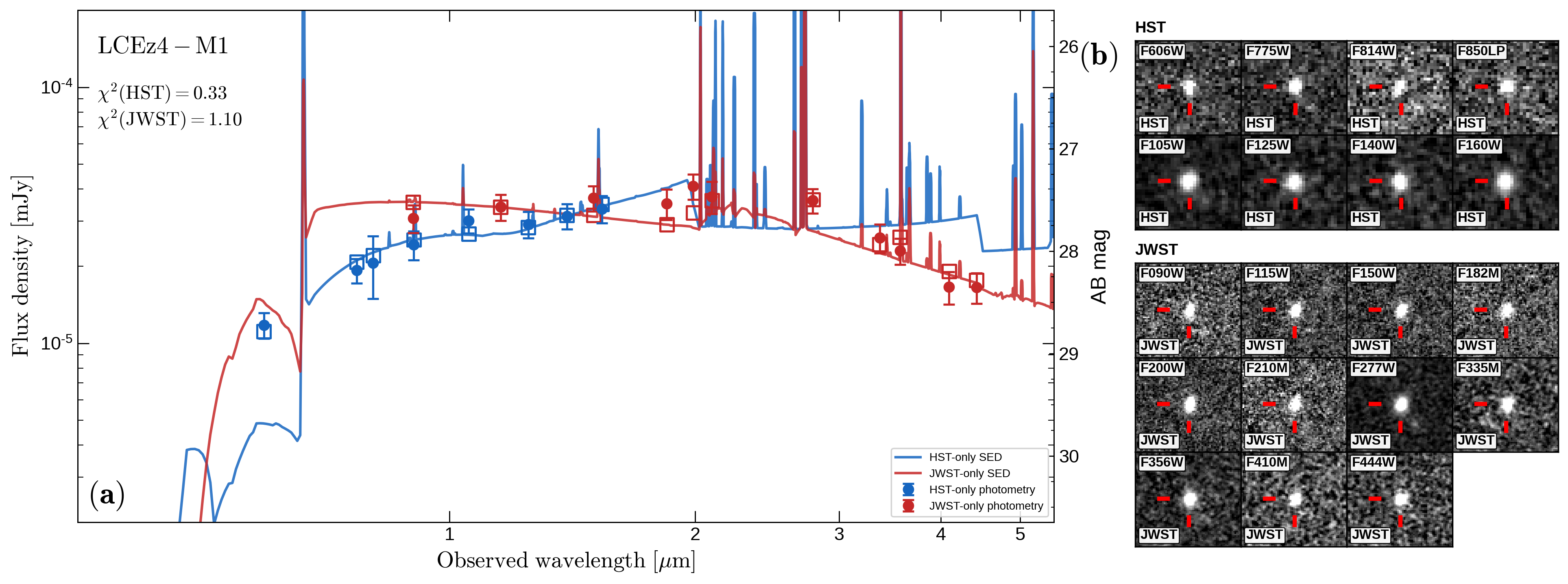}
    \caption{SED fitting and HST/JWST cutouts of LCEz4-M1. (a) Comparison between the SED fits obtained using the HST-only and JWST-only photometry. Filled symbols show the observed photometry, open squares show the model-predicted fluxes, and the blue and red curves show the best-fit HST-only and JWST-only model spectra, respectively. The displayed fit statistics are $\chi^2({\rm HST})=0.33$ and $\chi^2({\rm JWST})=1.10$. (b) HST and JWST cutouts centered on LCEz4-M1. The red ticks mark the source position. The sizes of each cutouts are $2\arcsec \times 2 \arcsec$.}
    \label{fig:sed-phot}
\end{figure*}

We first check the consistency of the photometry between the JADES DR2 and HLF catalogs. We perform SED fitting using the HLF and JADES photometry separately, and find that the two sets of photometry do not yield fully consistent SEDs (see Fig. \ref{fig:sed-phot}). A tension is present in the wavelength-overlap $\sim$0.8 to 1.6 $\mu$m region, with the HST bands being systematically lower than the JWST bands. The tension may arise either from the photometric method’s ability to accurately recover faint signals or from the intrinsic variability of LCEz4-M1. Including these HST photometric points drives the fit toward a redder UV continuum and a much younger, dustier, higher-SFR solution. Since the photometric data from JWST are much newer and with higher S/N, we therefore use only the JWST/NIRCam photometry to derive our fiducial physical properties, while retaining the HST-only fit as a systematic comparison. We also excluded NIRCam F460M and F480M from the JWST-only SED fitting due to the low S/N. The HST-only and JWST-only SED fits and the HST/JWST cutouts are shown in Figure~\ref{fig:sed-phot}.

The fitting procedure follows our previous work \citep{Yuan2021,zhu2024}, with updates to the stellar population model and SFH prescription. We use the BPASS v2.2 stellar population synthesis models \citep{Stanway2018}, including binary stellar evolution, with the Chabrier IMF \citep{chabrier2003} option extending to $300~M_{\odot}$. The star formation history is modeled using a delayed form with an optional recent exponential component. We allow the main stellar population age to vary from 2 to 50 Myr and the main SFH timescale to vary over $\tau_{\rm main}=5$, 10, and 20 Myr. The recent component age is fixed to 1 Myr, with recent-component mass fractions of 0, 0.001, 0.01, 0.02, and 0.1.

The stellar metallicity is allowed to vary over $Z=0.00001$, 0.0001, and 0.004. For dust attenuation, we assume the \citet{Calzetti2000} law for the stellar continuum, extended with the \citet{Leitherer2002} curve between the Lyman break and $1500~\text{\AA}$, with $E(B-V)_{\rm lines}=0.01$, 0.05, 0.1, 0.2, and 0.5. We set $E(B-V)_{\rm star}=E(B-V)_{\rm lines}$ and allow the attenuation-curve slope to vary over $\delta=-0.6$, $-0.4$, and $-0.2$. For nebular emission, we adopt the \citet{Cardelli1989} Milky Way extinction curve. The nebular emission is modeled using templates from \citet{Inoue2011} with $\log U$ fixed to $-2.5$, an electron density of $n_{\rm e}=100~{\rm cm^{-3}}$, and gas metallicities of $Z_{\rm gas}=0.0004$, 0.002, and 0.004. We also allow the nebular-module LyC escape fraction to vary over 0.3, 0.5, and 0.7, which controls the strength of nebular emission in the SED models.

The physical properties inferred from the JWST-only and HST-only SED fits are summarized in Table~\ref{tab: res}. We also tested a single young stellar population model, but the current flexible SFH prescription provides a better fit to the observed SED. The fiducial JWST-only fit gives an older and bluer solution, with an age of $45.93\pm16.08$ Myr, $\beta=-2.26\pm0.13$, and a moderate current SFR. By contrast, the HST-only fit gives a much younger and dustier starburst-like solution, with an age of $5.93\pm5.67$ Myr, $\beta=-1.44\pm0.18$, $\log({\rm SFR}/M_{\odot}\,{\rm yr^{-1}})=1.22\pm0.40$, and $\Sigma_{\rm SFR}=6.59~M_{\odot}\,{\rm yr^{-1}\,kpc^{-2}}$. Because of the ACS/JWST photometric tension and the higher spatial resolution and internal consistency of the JWST/NIRCam photometry, we adopt the JWST-only fit as the fiducial basis for the physical interpretation below. Using the fiducial JWST-only solution, we compare LCEz4-M1 with the observed $\beta$--$M_{\rm UV}$ relation at $z\sim7$ \citep{Bouwens2014} and with the star-forming main sequence at $z\simeq4.4$ \citep{Popesso2023}. LCEz4-M1 is consistent with the observed $\beta$-$M_{\rm UV}$ relation at $z \sim 7$. This behavior is similar to low-redshift LCEs, whose $\beta$-$M_{\rm UV}$ trend broadly follows the $z\sim7$ relation \citep{Chisholm2022}, and LCEz4-M1 also falls within the range occupied by high-redshift LyC leakers in GOODS-S \citep{zhu2024}. In the fiducial JWST-only fit, LCEz4-M1 is consistent with the star-forming main sequence within the uncertainties.

Using the effective radius derived in Section~\ref{sec:morp}, the fiducial JWST-only fit gives a star formation rate surface density of $\Sigma_{\rm SFR}=0.38~M_{\odot}\,{\rm yr^{-1}\,kpc^{-2}}$ for LCEz4-M1. This value is lower than those typically reported for the most compact low-$z$ LyC leakers \citep[e.g.,][]{Flury2022a}, and it is also below the very high-$\Sigma_{\rm SFR}$ regime emphasized in models of prodigious LyC leakers \citep{Cen2020}. Thus, under the fiducial JWST-only interpretation, LCEz4-M1 does not appear to be an extreme compact starburst. The HST-only fit would instead imply a compact starburst-like state, illustrating that this conclusion is sensitive to the HST/JWST photometric tension. Nevertheless, even the JWST-only $\Sigma_{\rm SFR}$ remains above the threshold associated with feedback-driven outflows \citep[$\sim0.1~M_{\odot}\,{\rm yr^{-1}\,kpc^{-2}}$;][]{Prusinski2021}.

We do not detect any prominent nebular emission lines in the JWST/NIRCam slitless spectra (F.\ Sun, priv.\ comm., PI of GO-7336). In particular, [O \textsc{iii}] $\lambda\lambda4959,5007$ is not detected in the F277W grism spectrum. The 3$\sigma$ detection limit is estimated by injecting Gaussian line profiles into the grism spectra, with the line flux and central wavelength randomly varied within a narrow range around the expected line wavelength. This procedure yields a 3$\sigma$ limit of line flux of $\sim 2.3\times10^{-18}~{\rm erg~s^{-1}~cm^{-2}}$.
The fiducial JWST-only broadband SED indicates that the flux excess attributable to these lines is weak, suggesting that the non-detections are likely caused by their intrinsic faintness. Using the fiducial JWST-only best-fit SED model, we estimate line fluxes of $f_{4959}=6.68\times10^{-19}\ \mathrm{erg\ s^{-1}\ cm^{-2}}$,
$f_{5007}=1.98\times10^{-18}\ \mathrm{erg\ s^{-1}\ cm^{-2}}$, which is slightly lower than the 3$\sigma$ detection limit.
These predicted fluxes imply that the current NIRCam slitless spectroscopic data may not be sufficiently deep to robustly detect these lines.

\subsection{Morphology and Environments}
\label{sec:morp}
\begin{figure*}[ht]
    \centering
    \includegraphics[width=0.7\linewidth]{fig3.png}
    \caption{Morphological analysis of LCEz4-M1 using the JWST/NIRCam F200W data. Each stamp has a size of $2\arcsec \times 2\arcsec$. The source is fitted both with a single-S\'ersic model and with a two-component S\'ersic model. The two rows show the fitting results for the single-component and two-component models, respectively. The reduced chi-squared values are similar for the two fits ($\chi^2_\nu \approx 0.5$).}
    \label{fig:morp}
\end{figure*}

The morphology of LCEz4-M1 is not clumpy on visual inspection. We further analyze its structure by fitting parametric surface-brightness models with \texttt{GALFIT} \citep{Peng2010}. Our morphological measurements are based on the JWST NIRCam F200W image. The point-spread function (PSF) is constructed using \texttt{PSFEx} \citep{Bertin2013}.

We fit a single-\sersic\ model and a two-component \sersic\ model with \texttt{GALFIT}. The best-fit model and the residual images are shown in Figure~\ref{fig:morp}. Both models reproduce the overall F200W light distribution of LCEz4-M1, with no obvious large-scale residuals.

To compare models with different numbers of free parameters, we adopt the Bayesian Information Criterion (BIC),$\mathrm{BIC} = \chi^{2} + k\,\ln N$,
where $k$ is the number of fitted parameters and $N$ is the number of data points used in the fit. We define $\Delta\mathrm{BIC}\equiv \mathrm{BIC}_{\rm double}-\mathrm{BIC}_{\rm single}$. We find $\Delta\mathrm{BIC} > 10$, strongly favoring the single-\sersic\ model over the two-component one.

LCEz4-M1 shows a compact morphology, with no clear evidence for an ongoing major merger. However, there is a faint companion close to the main source (see Figure~\ref{fig:spec-z}g).
If the companion candidate lies at the same redshift as LCEz4-M1, its projected separation is $\sim3.3$~kpc, which is within the expected virial radius of a $z\simeq4.4$ galaxy halo with $M_\star\sim10^8~M_\odot$ \citep[e.g.,][]{Barkana2001}.
This companion exhibits a strong flux excess in JWST/NIRCam F277W, suggesting that it may be a faint emission line galaxy. Because the companion candidate is undetected in most bands and has no secure MUSE or NIRCam grism redshift, its redshift and physical association with LCEz4-M1 remain uncertain.
Deeper and higher-resolution imaging and spectroscopy are required to determine its nature and relationship to the primary source.
The other labeled sources in Figure~\ref{fig:spec-z}e--g, MUSE 109498 and MUSE 110116, are foreground objects and are distinct from the companion candidate discussed here.

Although we do not identify clear merger signatures in the main galaxy, LCEz4-M1 may reside in a locally overdense environment. Using the JADES photometric-redshift catalog, we select galaxies within a redshift slice of $z = 4.444 \pm 0.1$ and find that LCEz4-M1 lies in a region with a relatively high projected surface density of galaxies. We also examine the MUSE-HUDF LAE catalog and select LAEs within a velocity window of $\Delta v \le 1000~\mathrm{km~s^{-1}}$ relative to LCEz4-M1. In this search, we identify 15 LAEs in the vicinity of the source. These nearby galaxies suggest that LCEz4-M1 may be located in a locally overdense region at similar redshift. However, given the depth of the available data and the lack of a completeness analysis, the current evidence is not sufficient to establish that the source resides in a proto-cluster. If LCEz4-M1 indeed lies in a dense environment, interactions and dynamical perturbations may influence the structure of the interstellar and circumgalactic medium and thus affect LyC escape.

\subsection{A Possible Post-burst LyC Escape Scenario}

The low Ly$\alpha$ EW and relatively weak rest-frame optical emission lines appear, at first sight, to be in tension with the high LyC escape fraction inferred for LCEz4-M1. In low-redshift LCE samples, Ly$\alpha$ EW, Ly$\alpha$ escape, $O_{32}$, and $\Sigma_{\rm SFR}$ have all been shown to be useful empirical tracers of LyC leakage, although with substantial scatter \citep{Flury2022a,Flury2022b,SaldanaLopez2022,Jaskot2024}. Similar positive trends between Ly$\alpha$ EW and escaping LyC flux have also been reported at $z\sim3$--4 \citep{Steidel2018,Pahl2021}. However, these relations are not expected to remain strictly monotonic when $f_{\rm esc}$ approaches unity. In this regime, a large fraction of ionizing photons escapes before being absorbed by the ISM, reducing the photon budget available to power recombination emission. Therefore, Ly$\alpha$ emission can be suppressed even when the LyC escape fraction is high \citep[e.g.,][]{Dijkstra2014}. By the same argument, non-resonant recombination lines such as H$\alpha$ and H$\beta$ can also be weakened.

A time-dependent escape scenario provides a natural explanation for this behavior in the fiducial JWST-only interpretation. Radiation-hydrodynamic simulations predict that LyC escape can lag behind the peak of star formation by several to $\sim10$ Myr, because young massive stars are initially embedded in dense birth clouds and efficient escape begins only after stellar feedback and supernovae have opened low-column-density channels \citep{Kimm2014,Ma2015,Trebitsch2017}. In this picture, a galaxy can show high $f_{\rm esc}$ after the instantaneous SFR, nebular-line EWs, and global $\Sigma_{\rm SFR}$ have already declined. This is consistent with the ``Remnant Leaker'' phase proposed by \citet{Katz2023}, in which galaxies maintain high LyC escape even when ${\rm SFR}_{10\,{\rm Myr}}<{\rm SFR}_{100\,{\rm Myr}}$; the fiducial JWST-only fit for LCEz4-M1 shows the same sense of recent decline, with ${\rm SFR}_{10\,{\rm Myr}}/{\rm SFR}_{100\,{\rm Myr}}\simeq0.16$. Binary stellar evolution can further extend the production of ionizing photons after the main burst, enhancing the relevance of such delayed escape phases \citep{Secunda2020}. Thus, under the fiducial JWST-only solution, the combination of high inferred $f_{\rm esc}$, low Ly$\alpha$ EW, relatively weak H$\beta$+[O\,{\sc iii}], and moderate current $\Sigma_{\rm SFR}$ may indicate that LCEz4-M1 is observed in a post-burst or geometry-dominated LyC-leaking phase, rather than as a classical extreme emission-line compact starburst.

We note that \citet{Goovaerts2026} interpret the same source as undergoing a very recent burst of star formation, and our HST-only fit would favor a similarly young, high-SFR solution. The post-burst interpretation therefore depends on adopting the fiducial JWST-only photometry, and should be regarded as one plausible scenario rather than the only allowed SFH. Nevertheless, both the JWST-only and HST-only interpretations imply a time-dependent LyC escape scenario in which bursty star formation and ISM geometry play a central role.

\section{Summary}
We report LCEz4-M1 as a candidate LCE at $z = 4.444$, among the highest-redshift LCE candidates currently known. Using deep VLT/MUSE spectroscopy together with extensive HST and JWST images and photometry, we confirm the redshift of the source and the robustness of the LyC detection. We further characterize LCEz4-M1 by estimating its LyC escape fraction, measuring its
physical properties, and quantifying its morphology.

The redshift of LCEz4-M1 is determined from its Ly$\alpha$ emission line, which shows an asymmetric, red-skewed line profile. We combine the high-resolution HST imaging with the MUSE data cube to verify that the Ly$\alpha$ emission is spatially associated with the galaxy rather than from any contaminating sources. The LyC signal is detected independently in two datasets, with ${\rm S/N} \sim3.7$ in the HST ACS/WFC F435W image and ${\rm S/N} \sim 2.8-3.0$ in the MUSE data.

Based on the HST/ACS F435W photometry and the MUSE spectroscopy, and adopting the maximum IGM transmission allowed by the \citet{Steidel2018} prescription, we estimate conservative lower-limit LyC escape fractions of $f_{\rm esc}=0.82^{+0.13}_{-0.17}$ and $0.75^{+0.18}_{-0.28}$, respectively.

We further analyze LCEz4-M1's physical properties, morphology. In the fiducial JWST-only fit, LCEz4-M1 is compact, but its current galaxy-integrated star formation rate surface density, $\Sigma_{\rm SFR}=0.38~M_{\odot}\,{\rm yr^{-1}\,kpc^{-2}}$, is lower than those typically found in the most compact low-redshift LyC leakers. Its UV continuum slope and luminosity are consistent with the observed $\beta$--$M_{\rm UV}$ relation at $z\sim7$, and it is consistent with the star-forming main sequence within the uncertainties. An HST-only fit instead gives a younger, dustier, high-SFR solution, illustrating the sensitivity of the inferred SFH to the HST/JWST photometric tension. We do not find clear morphological signatures of an ongoing merger, although the presence of a faint companion at $\sim0\farcs5$ away suggests a possible minor interaction. We also find that the source may lie in a locally overdense environment.

Under the fiducial JWST-only interpretation, the combination of high inferred $f_{\rm esc}$, low Ly$\alpha$ EW, relatively weak H$\beta$+[O\,{\sc iii}], and moderate current $\Sigma_{\rm SFR}$ may indicate a post-burst or geometry-dominated LyC-leaking phase rather than a classical extreme emission-line compact starburst.

LCEz4-M1 is among the few LyC detections reported at $z>4$, providing a valuable laboratory to investigate the physical conditions that allow LyC escape. Future deep spectroscopy would help assess whether the special properties of the galaxy or its environment facilitate the escape of such a large amount of LyC photons, making them visible to us despite the extremely high redshift. In addition, the upcoming Chinese Space Station Survey Telescope (CSST) main survey \citep{2026SCPMA..6939501C} and Multi-Channel Imager’s imaging survey \citep{Zheng2026RAA} will deliver wide-area and deep multi-band data, which will be essential to build larger samples of similar systems. The CSST main survey will cover $\sim17,500~{\rm deg}^2$ in the wide-field survey and $\sim400~{\rm deg}^2$ in the deep survey, reaching 5$\sigma$ depths of 25.4 and 26.7 mag in both the NUV and $u$ bands, respectively. These bands can probe redshifted LyC emission at $z\sim1.8$--3.4. The MCI survey will reach $\sim30$ mag in the $u$ band over an area several times larger than the Hubble eXtreme Deep Field \citep{Illingworth2013}, with medium- and narrow-band filters that can help isolate LyC emission from the non-ionizing UV continuum.

\begin{acknowledgments}
We thank the anonymous referee for the very helpful suggestion, which improved the manuscript.
We thank Dr. Fengwu Sun for providing the JWST/NIRCam grism data from program GO-7336 and for helpful discussions.

This work is supported by the National Key R\&D Program of China No.2022YFF0503402. We also acknowledge the science research grants from the China Manned Space Project, especially, NO.  CMS-CSST-2025-A18. ZYZ acknowledges the supports by the Shanghai Leading Talent Program of Eastern Talent Plan (LJ2025051) and the China-Chile Joint Research Fund.

This work is based on observations made with the NASA/ESA/CSA James Webb Space Telescope. The data were obtained from the Mikulski Archive for Space Telescopes at the Space Telescope Science Institute, which is operated by the Association of Universities for Research in Astronomy, Inc., under NASA contract NAS 5-03127 for JWST.

This work is based on observations from the MUSE Extreme Deep Field \citep{Bacon2023}, obtained from the ESO Science Archive Facility \dataset[DOI: 10.18727/archive/85]{https://doi.org/10.18727/archive/85}. The other data used in this paper can be found in MAST: \dataset[DOI: 10.17909/T91019]{https://doi.org/10.17909/T91019} \citep{Illingworth2015_HLF}, \dataset[DOI: 10.17909/8tdj-8n28]{https://doi.org/10.17909/8tdj-8n28}, and \dataset[DOI: 10.17909/fsc4-dt61]{https://doi.org/10.17909/fsc4-dt61}.

\end{acknowledgments}

\begin{contribution}
S. Zhu led the data analysis, including the photometry, escape fraction estimation, SED fitting, and morphological analysis, and wrote the manuscript. Z.-Y. Zheng conceived and supervised the project, contributed to the scientific interpretation, and revised the manuscript. F. Bian, F.-T. Yuan, C. Jiang, X. Zhang, R. Lin, and Y. Guo contributed to the interpretation and manuscript revision.
\end{contribution}

\facilities{HST (ACS, WFC3), VLT (IASSC, MUSE), Spizter (IRAC), JWST (NIRCam)}
\software{astropy \citep{2013A&A...558A..33A,2018AJ....156..123A,2022ApJ...935..167A},  Source Extractor \citep{1996A&AS..117..393B}, PSFEx \citep{Bertin2013}, CIGALE \citep{Boquien2019}, GALFIT \citep{Peng2010}, Numpy \citep{harris2020array}, Matplotlib \citep{Hunter:2007}, Scipy \citep{2020SciPy-NMeth}}

\newpage
\bibliography{refs}{}
\bibliographystyle{aasjournalv7}
\end{document}